\documentclass[aps,prev,twocolumn,preprintnumbers,floatfix,nofootinbib]{revtex4-1}
\pdfoutput=1
\usepackage{graphicx}
\usepackage{bm}
\usepackage{times}
\usepackage{hyperref}
\usepackage{slashed}
\usepackage{color}
\usepackage{appendix}
\usepackage{mathtools}
\usepackage{epsfig}
\usepackage{dcolumn}
\usepackage{textcomp}

\begin{document} 

\title{Roads for Right-handed Neutrino Dark Matter: Fast Expansion, Standard Freeze-out, and Early Matter Domination}

\author{Giorgio Arcadi$^{1,2,3}$}
\email{giorgio.arcadi@uniroma3.it}
\author{Jacinto Paulo Neto$^{4,5}$}
\email{jacinto.neto.100@ufrn.edu.br}
\author{Farinaldo S. Queiroz$^{4,5,6}$}
\email{farinaldo.queiroz@iip.ufrn.br}
\author{Clarissa Siqueira$^{7}$}
\email{csiqueira@ifsc.usp.br}
\affiliation{$^1$ Dipartimento di Matematica e Fisica, Universita di Roma Tre, Via della Vasca Navale 84, 00146, Roma, Italy}
\affiliation{$^2$ INFN Sezione di Roma Tre, Via della Vasca Navale 84, 00146, Roma, Italy}
\affiliation{$^3$ Dipartimento di Scienze Matematiche e Informatiche, Scienze Fisiche e Scienze della Terra, Universita degli Studi di Messina, Via Ferdinando Stagno d'Alcontres 31, I-98166 Messina, Italy}
\affiliation{$^4$ Departamento de F\'isica, Universidade Federal do Rio Grande do Norte, 59078-970, Natal, RN, Brasil}
\affiliation{$^5$ International Institute of Physics, Universidade Federal do Rio Grande do Norte, Campus Universitário, Lagoa Nova, Natal-RN 59078-970, Brazil}
\affiliation{$^6$ Millennium Institute for Subatomic Physics at High-Energy Frontier (SAPHIR),
Fernandez Concha 700, 7550196, Santiago, Chile}
\affiliation{$^7$ Instituto de F\'isica de S\~ao Carlos, Universidade de S\~ao Paulo, Avenida Trabalhador S\~ao-carlense, 400, S\~ao Carlos-SP, 13566-590, Brasil.}

\begin{abstract}
\noindent
Right-handed neutrinos appear in several extensions beyond the Standard Model, specially in connection to neutrino masses. Motivated by this, we present a model of right-handed neutrino dark matter that interacts with Standard Model particles through a new gauge symmetry as well as via mass mixing between the new vector field and the Z boson, and investigate different production mechanisms. We derive the dark matter relic density when the Hubble rate is faster than usual, when dark matter decouples in a matter domination epoch, and when it decouples in a radiation domination regime, which is then followed by a matter domination era. The direct detection rate features a spin-independent but velocity suppressed operators, as well as a spin-dependent operator when the mass mixing is correctly accounted for. We put all these results into perspective with existing flavor physics, atomic parity violation, and collider bounds. Lastly, we outline the region of parameter space in which a weak scale right-handed neutrino dark matter stands as a viable dark matter candidate. 

\end{abstract}


\maketitle
\flushbottom

\section{\label{Intro} Introduction}

The presence of dark matter in our universe is ascertained through a variety of dataset, going from dwarf galaxy scales to cosmological scales. The precision acquired by Cosmic Microwave Background (CMB) probes allowed us to quantify precisely the abundance of dark matter in our Universe. It accounts for nearly 27\% of the total energy density \cite{Planck:2018vyg}. Its nature is unknown, though. Massive particles that feature weak interactions with Standard Model (SM) particles dubbed (WIMPs) have stood out from the crowd, driving most experimental efforts for many years \cite{Arcadi:2017kky}. Currently, other plausible dark matter candidates such as the axion and axion-like particles, which have been always theoretically compelling, gained lots of interest from the community  due to the non-positive signals in WIMP searches \cite{Leane:2018kjk}. However, no solid positive signals have been observed, favoring other dark matter candidates either. The customary assumption that dark matter had a thermal production have given us a misapprehension that a positive signal should appear in the current generation of experiments. There are ways to successfully have a thermal dark matter candidate while yielding no signal at current direct detection experiments. Hence, this sudden disbelief for the WIMP miracle does not seem justified. It has always been worthwhile to explore well-motivated dark matter production mechanisms that lead to attainable detection, specially if the mechanism still leaves imprints at current experiments.

Besides the dark matter siege, neutrino masses stand as a concrete evidence for physics beyond the Standard Model. With the observation of neutrino oscillations,  we have concluded that at least two neutrinos are massive \cite{Super-Kamiokande:2001bfk,SNO:2002tuh,KamLAND:2008dgz,T2K:2011ypd,DoubleChooz:2011ymz,DayaBay:2012fng,RENO:2012mkc,MINOS:2013xrl}. It should be noted that the sum of the neutrino masses is constrained by CMB observations because the energy density of neutrinos which depends on the sum of neutrino masses affects both the relativistic and non-relativistic energy density relevant to the derivation of the CMB temperature and polarization power spectra  \cite{Abazajian:2013oma}.  In fact, CMB probes give rise to the most stringent bound that reads $\Sigma m_\nu < 0.12$~eV \cite{2020}. Although, the individual masses are unknown. 

An elegant solution to theoretically generate neutrino masses is offered by right-handed neutrinos. In particular, if copies of right-handed neutrinos are added to the Standard Model,  one can naturally address neutrino masses through a type I seesaw mechanism when both Dirac and Majorana mass terms are included in the Yukawa Lagrangian. The Majorana mass is expected to be very large to play the seesaw and bring the active neutrino masses down below the eV scale. 

That said, one may wonder if the mechanisms behind the presence of dark matter in our universe is somehow connected to neutrino masses.  This question has driven a multitude of works. In particular, we will be interested in connecting these two new physics landmarks in the context of the type I seesaw mechanism \cite{Aoki:2008av,Nomura:2017wxf,Nomura:2017jxb,Cai:2018upp,Gehrlein:2019iwl,VanLoi:2019eax,Singirala:2019moc,DiBari:2019amk,Mishra:2019gsr,Chao:2012mx,Restrepo:2019soi,Das:2019pua,Jaramillo:2020dde,Dror:2020jzy,VanDong:2020cjf}, which features the Lagrangian,

\begin{equation}
    \mathcal{L} \supset y_{ab} \overline{L_a}\, \Phi N_{bR} + \frac{M_a}{2} \overline{N_{aR}^c} N_{aR},
    \label{eq1}
\end{equation}where $\Phi$ is the SM Higgs doublet, and $M$ the Majorana mass term for the right-handed neutrinos. Note that this Majorana mass is a $3\times3$ matrix. After spontaneous electroweak symmetry breaking, the first term yields a Dirac mass term for the active neutrinos. The presence of a Majorana mass allow us to employ the so-called type I seesaw mechanism that results into $m_\nu \simeq m_D^T M^{-1} m_D$ and $m_N\sim M$.

For Yukawa couplings of order one in Eq.~\eqref{eq1}, this simple and plausible framework requires the right-handed neutrino masses to be very large, above $10^{12}$~GeV. Consequently, this mechanism is hardly testable by existing experiments.  It would be interesting if one could successfully explain neutrino masses and dark matter without invoking a very large energy scale suppression. A possible route is to consider light right-handed neutrinos with masses around $1$~keV. If one of the three right-handed neutrinos is lighter than the others, then it can be made cosmologically stable as long as the mixing between active and right-handed neutrinos is sufficiently small. This scenario is dubbed sterile neutrino dark matter\footnote{For sterile neutrino and Higgs portal considering freeze-in production mechanism, see \cite{De_Romeri_2020}.} \cite{Dodelson:1993je,Asaka:2005an}.  The dark matter relic density is found through a non-resonant production \cite{Dodelson:1993je}, which is based on the active-sterile neutrino oscillations. The  approximate relic density is \cite{Dolgov:2000ew,Abazajian:2001nj,Abazajian:2005gj,Asaka:2006nq,Kusenko:2009up}, 

\begin{equation}
\Omega_{N}h^2 \sim 0.1 \left(\frac{\sin^2\theta_i}{3 \times 10^{-9}}\right)\left(\frac{m_N}{3~\mathrm{keV}}\right)^{1.8} , 
\label{eq3}
\end{equation}
where $\sin^2\theta_i \sim \sum_{a} y_{ab}^2 v^2/M^2$ is the active-sterile neutrino mixing and $v$ is the SM Higgs vacuum expectation value. As mentioned earlier, the sterile neutrino is unstable because of its mixing with the active neutrinos. In fact, this mixing leads to the decays $N\rightarrow \nu\nu\nu$, and $N \rightarrow \nu \gamma$ decay \cite{Queiroz:2014yna,Garcia-Cely:2017oco}. The latter decay has been extensively looked for in X-rays surveys. However, such X-ray searches combined with structure formation requirements rule out this sterile neutrino production  \cite{Abazajian:2017tcc,Boyarsky:2018tvu}. There are ways to bypass this exclusion by evoking other production mechanisms \cite{Shi:1998km,Schneider:2016uqi}. Instead of dwelling on light sterile neutrinos as dark matter, we investigate under which circumstance we can accommodate the type I seesaw mechanism, while hosting a TeV scale right-handed neutrino dark matter. Note, we have renamed sterile neutrino to right-handed neutrino. Theoretically speaking, they are the same field, but when ones refers to sterile neutrino typically small masses and suppressed mixing angles are adopted. When this sterile neutrino is heavy, with masses above $1$~MeV, the term right-handed neutrino is used more often\footnote{We can also find heavy right-handed neutrinos in asymmetric dark matter scenario \cite{Ahmadvand_2021,Ahmadvand:2021vxs}}. 

Anyway, how can we have a weak scale right-handed neutrino dark matter having the presence of the first term in Eq.~\ref{eq1}, which is necessary for the type-I seesaw to work but make it unstable? The idea is to invoke a $\mathcal{Z}_2$ symmetry, where only one of the right-handed neutrinos is odd under it \cite{Okada:2010wd,Okada:2016tci,Okada:2016gsh,Okada:2018ktp}. In this way, two massive right-handed neutrinos will play a role in the seesaw mechanism, whereas the lightest becomes a plausible dark matter candidate. Notice that in general, one needs to either invoke fine-tuning in the yukawa couplings to make the right-handed neutrino a viable dark matter candidate or symmetries. We focus on the latter. Similar works have been carried out in this direction \cite{Campos:2017dgc,Camargo:2018klg,Blasi:2020wpy}. Although, our work extends previous studies because we go beyond the standard freeze-out case.  Concretely, we investigate the right-handed neutrino dark matter scenario under three different cosmological histories, namely fast expansion, matter-domination during freeze-out, and matter-domination after freeze-out. These cosmological histories change the theoretical predictions for the dark matter relic density compared to the standard freeze-out case, and consequently lead to different conclusions. Besides computing the dark matter relic density under different cosmological scenarios, we also compute the dark matter scattering cross-section. We highlight that for right-handed neutrino dark matter, the dark matter scattering cross-section does not fall into the standard classification of spin-independent or spin-dependent dark matter-nucleon scattering. Therefore, we need to compare the resulting scattering rate with the data to derive the correct limits. Lastly, we also include collider bounds on the mediator masses based on the LHC search for heavy dilepton resonances.

All this procedure is done in a well motivated Two Higgs Doublet Model (2HDM) featuring an Abelian gauge symmetry. A model that can elegantly solve the flavor problem present in general 2HDM constructions, and address neutrino masses via the type I seesaw mechanism. 2HDM featuring new gauge symmetries are getting increasingly interest from the community because it offers new collider signatures \cite{Huang:2017bto,Accomando:2017qcs,Camargo:2018klg,Arhrib:2018sbz,Huang:2019obt}, and avenues to be explored concerning atomic parity violation \cite{Davoudiasl:2012ag,Davoudiasl:2012qa}, neutrino-electron scattering \cite{Lindner:2018kjo}, axion-like particles \cite{Dias:2021lmf}, exotic Higgs decays \cite{Ko:2013zsa,Chen:2018wjl,Camargo_2020}, neutrino masses \cite{Heeck:2011wj,Camargo:2018uzw,Cogollo:2019mbd}, flavor studies \cite{Lindner:2016bgg,Botella:2018gzy,Ordell:2019zws,Ferreira:2020ana}, and dark matter \cite{Ko:2014uka,Nomura:2017wxf,Camargo:2019ukv,Chen:2019pnt,Lindner:2020kko,Arcadi:2020aot,Dirgantara:2020lqy}, among others \cite{DelleRose:2017xil,DelleRose:2018eic,DelleRose:2019ukt,DelleRose:2020oaa,Arcadi:2021yyr}. Hence, it is definitely worthwhile to investigate how feasible is to host a dark matter candidate connected to neutrino masses and cosmological histories that go beyond the standard freeze-out.

\section{\label{model}The model}

2HDMs have been extensively studied in the literature \cite{Lee:1973iz,Haber:1984rc,Branco:2011iw} for naturally keeping the $\rho$ parameter unchanged despite the extended scalar sector. Furthermore, it leads to interesting collider searches that are within reach of the LHC. Although, the canonical version of the 2HDM suffers from flavor changing neutral interactions and does not address neutrino masses. It would be appealing if one could solve both issues via gauge symmetries. This is precisely the idea behind the 2HDM-U(1). An Abelian gauge symmetry is incorporated to the 2HDM in such a way that only one scalar doublet contribute to fermion masses, and three right-handed neutrinos  are required to cancel the gauge anomalies and consequently play a role in the type I seesaw mechanism aforementioned.  We will choose the new Abelian gauge symmetry to be the Baryon-Lepton (B-L) number, for concreteness, but we highlight that other gauge symmetries are also conceivable. The purpose of our work is to incorporate dark matter without adding new fields and explore the impact of our conclusions under different cosmological histories. Concerning the model itself, which is based on the $SU(3)_c \times SU(2)_L \times  U(1)_Y \times U(1)_{B-L}$, the fermion content features, 

\begin{equation}
\begin{split}
Q _{aL} = \begin{pmatrix} u _{aL} \\ d_{aL}\end{pmatrix} &\sim ({\bf 3},{\bf 2},1/6,1/3), \nonumber \\
u_{aR} \sim ({\bf 3},{\bf 1},2/3,1/3) \,\, &\mathrm{and} \,\, d_{aR} \sim ({\bf 3},{\bf 1},-1/3,1/3),
\end{split}
\end{equation}and,
 
\begin{equation}
\begin{split}
L _{aL} = \begin{pmatrix} e _{aL} \\ \nu_{aL}\end{pmatrix} &\sim ({\bf 1},{\bf 2},-1/2,-1),\nonumber \\
e_{aR} \sim ({\bf 1},{\bf 1},-1,-1) \,\, &\mathrm{and} \,\, N_{aR} \sim ({\bf 1},{\bf 1},0,-1).
\end{split}
\end{equation}

Notice that thus far the only difference to the SM is the presence of right-handed neutrinos. The model has three scalar fields,

\begin{equation}
\Phi _1 = \begin{pmatrix} \phi ^+ _1 \\ \phi^0_1\end{pmatrix} \sim ({\bf 1},{\bf 2},1/2,2),
\end{equation}
\begin{equation}
\Phi _2 = \begin{pmatrix} \phi ^+ _2 \\ \phi^0_2\end{pmatrix} \sim ({\bf 1},{\bf 2},1/2,0),
\end{equation}
and
\begin{equation*}
\Phi _s  \sim ({\bf 1},{\bf 1},0,2).
\end{equation*}
where the subscript $a=1,2,3$ accounts for the three generations. We stress that the presence of the scalar singlet is twofold: (i) it breaks the U(1) gauge symmetry generating mass to the $Z^\prime$ gauge boson; (ii) it gives rise to the Majorana mass for the right-handed neutrinos. In the original version of the type I seesaw, we have a Majorana bare mass term for the right-handed neutrino masses, but in our work it is generated through a mechanism of spontaneous symmetry breaking. This fact will be clear below. Fermions get mass through the Yukawa Lagrangian $\mathcal{L_Y} = \mathcal{L}_{Y_1}+ \mathcal{L}_{Y_2}$, where,

\begin{equation}
\mathcal{L}_{Y_1} = -y^d_{ab} \bar{Q} _a \Phi _2 d_{bR} - y_{ab}^u \bar{Q}_a \widetilde \Phi _2 u_{bR} - y_{ab}^e \bar{L}_a \Phi _2 e_{bR} + h.c.,
\label{eq4}
\end{equation}and,
\begin{equation}
\begin{split}
\mathcal{L}_{Y_2} &\supset  -y_{ab} \bar{L}_a \widetilde \Phi _2 N_{bR} - y^{M}_{ab}\overline{(N_{aR})^{c}}\Phi_{s}N_{bR}.
\label{eq5}
\end{split}
\end{equation}

Having a weak scale right-handed neutrino from Eq.~\eqref{eq5} requires the addition of an extra discrete symmetry. We will invoke a $\mathcal{Z}_2$ symmetry to stabilize, say $N_{1R}$, the lightest right-handed neutrino. In this way, the first term in Eq.~\eqref{eq5} involves only two right-handed neutrinos, whereas the latter remains unaltered as long as $y^M_{ij}$ is diagonal. The two right-handed neutrinos in the Dirac mass term act the type I seesaw mechanism leading to two massive active neutrinos and a massless one, in agreement with neutrino oscillation observations \cite{Tanabashi:2018oca}. This is a well-known fact which can be directly seen using the Casas-Ibarra parametrization where the neutrino masses are directly tied to the Yukawa couplings. As we are basically removing one Yukawa coupling, the first term of Eq.\eqref{eq5} which is 6x6 matrix, should have at least four non-zero entries. Using the Casas\&Ibarra parametrization one can see that the Yukawa matrix will depend on two right-handed neutrinos masses and two active neutrinos masses \cite{Ibarra:2005qi}. We emphasize we have assumed that the Majorana mass matrix is purely diagonal.

The scalar potential, in agreement with the gauge charges of the doublets, including the singlet scalar, is given by,
\begin{equation}
\begin{split}
V \left( \Phi_{1,\, 2 ,\, s}\right) &= m_{11} ^2 \Phi _1 ^\dagger \Phi _1 + m_{22} ^2 \Phi _2 ^\dagger \Phi _2 + m_s ^2 \Phi _s ^\dagger \Phi _s \\ 
&+ \frac{\lambda _1}{2} \left( \Phi _1 ^\dagger \Phi _1 \right) ^2 + \frac{\lambda _2}{2} \left( \Phi _2 ^\dagger \Phi _2 \right) ^2 + \frac{\lambda _s}{2} \left( \Phi _s ^\dagger \Phi _s \right) ^2  \\
&+ \lambda _3 \left( \Phi _1 ^\dagger \Phi _1 \right) \left( \Phi _2 ^\dagger \Phi _2 \right) + \lambda _4 \left( \Phi _1 ^\dagger \Phi _2 \right) \left( \Phi _2 ^\dagger \Phi _1 \right) \\
&+ \mu _1 \Phi _1 ^\dagger \Phi _1 \Phi _s ^\dagger \Phi _s + \mu _2 \Phi _2 ^\dagger \Phi _2 \Phi _s ^\dagger \Phi _s \\
&+ \left( \mu \Phi _1 ^\dagger \Phi _2 \Phi _s + h.c. \right).
\end{split}
\label{eq6}
\end{equation}

The neutral components of all scalars acquire vacuum expectation values (\textit{vev}) as
\begin{equation}
\Phi _i =  \frac{1}{\sqrt{2}}\left( v_i + \rho _i + i\eta _i \right),
\end{equation}
and
\begin{equation*}
\Phi _s = \frac{1}{\sqrt{2}} \left( v_s + \rho _s + i \eta _s \right),
\end{equation*}
which give mass to all fermions and gauge bosons. It is important to emphasize that the dark matter candidate $N_{1R}$ becomes massive through the singlet scalar \textit{vev}, $v_s$,
\begin{equation}
m_{N} = \frac{y_N^M}{\sqrt{2}}v_s.    
\end{equation}
Concerning the gauge sector, kinetic mixing arises at tree level,
\begin{equation}
\mathcal{L} _{\rm gauge} =  - \frac{1}{4} \hat{B} _{\mu \nu} \hat{B} ^{\mu \nu} + \frac{\epsilon}{2\, cos \theta_W} \hat{X} _{\mu \nu} \hat{B} ^{\mu \nu} - \frac{1}{4} \hat{X} _{\mu \nu} \hat{X} ^{\mu \nu},
\label{Lgaugemix1}
\end{equation}
with $\epsilon$, the kinetic mixing parameter, $\theta_W$, the Weinberg angle. The Lagrangian responsible for gauge bosons' masses is given by
\begin{equation}
\mathcal{L}=(D^\mu \phi_1)^\dagger (D_\mu \phi_1)+ (D^\mu \phi_2)^\dagger (D_\mu \phi_2)+(D^\mu \phi_s)^\dagger (D_\mu \phi_s) 
\label{lag:gauge}
\end{equation}
where the covariant derivative is,
\begin{equation}
D_\mu = \partial _\mu + ig T^a W_\mu ^a + ig^\prime \frac{Q_{Y}}{2} \hat{B}_\mu + ig_X \frac{Q_X}{2} \hat{X}_\mu,
\label{Dcovgeral}
\end{equation}
with $Q_Y$ and $Q_X$ the charges under Yukawa and $B-L$ gauge symmetries, respectively, and $g$ and $g^\prime$ the usual gauge couplings associated with $SU(2)_L$ and $U(1)_Y$ symmetries, respectively, and $g_X$, the $B-L$ coupling. Diagonalizing the matrices for $X_\mu$ and $Z_\mu$ bosons, for details we recommend \cite{Camargo:2019ukv}, we get the following gauge boson masses
\begin{eqnarray}
m^{2}_{W} &=& \frac{1}{4} g^2 v^2, \\
m^{2}_{Z} &=& \frac{1}{4} g_Z^2 v^2, \\
m^{2}_{Z^\prime} &=& g_X^2 v_s^2 +g_X^2 v^2 \cos^2{\beta} \sin^2{\beta}, 
\end{eqnarray}
where $v^2 = v_1^2 + v_2^2 = 246^2$~GeV$^2$ and $g_Z = g/\cos{\theta_W}$. The Lagrangian in Eq.~\eqref{lag:gauge} is also responsible for the interactions between gauge bosons and scalars, for example $Z^\prime W ^+ W^-$, $Z^\prime W^+ H^-$, $H Z^\prime Z^\prime$, $h Z^\prime Z^\prime$, $H Z Z^\prime$, which are also relevant for the dark matter phenomenology. These couplings can be found in \cite{Camargo:2019ukv}. We emphasize that we have taken the kinetic mixing to be zero at tree-level. At one-loop level, it can be safely ignored, as our phenomenology will be driven by the gauge coupling $g_X$.

The dark matter phenomenology of the model is governed by the neutral current involving the Z and $Z^\prime$ gauge bosons that reads \cite{Klasen:2016qux,Campos:2017dgc},
\begin{equation}
\begin{split}
\mathcal{L_{\rm NC}} = &- e J ^\mu _{ em} A_\mu - \frac{g_Z}{2\cos\theta_W} J ^\mu _{NC} Z_\mu \nonumber \\
&-\left( \epsilon e J^\mu _{em} +  \frac{\epsilon _Z g}{2\cos\theta_W}  J^\mu _{NC} \right) Z' _\mu \\
&- \frac{g_X}{2} Q_{X_f} \left( \bar{\psi} _f \gamma ^\mu \psi _f \right) Z^\prime _\mu \\
&+\frac{1}{4} g_X \left(N_{1R} \gamma^\mu \gamma_5 N_{1R}\right) Z^{\prime}_\mu, \nonumber
\end{split}
\label{zzgeralcoma1}
\end{equation}where $Q_{X_f}=-1$ for charged leptons, $Q_{X_f}=1/3$ for quarks, and with,
\begin{equation}
 \epsilon_Z \equiv \dfrac{2 g_{X}}{g_{Z}}\cos^{2}\beta,
\end{equation}
and,
\begin{equation}
    J^\mu_{NC} = \left( T_{3f} - 2Q_{Yf} \sin^2\theta_W   \right) \bar{\psi}_f \gamma^\mu \psi_f - T_{3f} \bar{\psi}_f \gamma^\mu \gamma_5 \psi_f . 
\end{equation}

Identifying $N_{1R}$ as our dark matter candidate, we can straightforwardly carry out the dark matter phenomenology within the standard freeze-out and $\Lambda$CDM model. This scenario was investigated somewhere else \cite{Camargo:2019ukv}. Our plan is to go beyond and explore other cosmological scenarios. To do so, we start reviewing the important ingredients of early dark energy-like, early radiation, and early matter domination epochs. We have now set the particle physics model, we will review the cosmological background.

%%%%%%%%%%%%%%%%%%%%%%%%%%%%%%%%%%%%%%%%%%%%%%%%%%%%%%%%%%%%%%%%%%%%%%%%%%
%%%%%%%%%%%%%%%%%%%% Non-standard Cosmology Histories %%%%%%%%%%%%%%%%%%%%
%%%%%%%%%%%%%%%%%%%%%%%%%%%%%%%%%%%%%%%%%%%%%%%%%%%%%%%%%%%%%%%%%%%%%%%%%%
\section{\label{nstdcosmologies} Non-standard Cosmology Histories}

In this section, we explore how the dark matter relic density may be affected by a non-standard cosmological history, where the right-handed neutrino dark matter candidate decouples from thermal equilibrium in different scenarios. The key aspect of different cosmology histories is the effect on the Hubble expansion rate, because it directly impacts the Boltzmann equations.
The Hubble rate evolves differently for matter, radiation and dark energy-like fields. Thus,  depending on which component dominates $H$,  distinct solutions to the dark matter relic abundance are found.

The standard cosmology predicts that the Universe is radiation dominated at early times. However, before the Big Bang Nucleosynthesis (BBN), we have freedom to choose different cosmological histories for short period of times. For example, the expansion rate may be governed by matter or a new field, which leads to a faster expansion rate. That said, we tease out the dark matter phenomenology when the dark matter particles decouples during: (i) an expansion rate faster than radiation (Sec. \ref{fastertheory}); (ii) a radiation-dominated era (ERD) followed by a matter-dominated period (Sec.\ref{ERDtheory}); and (iii) in a matter-dominated phase (EMD) (Sec. \ref{EMDtheory}). We start reviewing the key ingredients of a faster expansion episode and how it changes the Boltzmann equation.

%%%%%%%%%%%%% Faster Than Usual Early Expansion %%%%%%%%%%%%%%%%%%
\subsection{\label{fastertheory}Faster Than Usual Early Expansion}

Assuming that a new scalar field $\phi$ has an energy density that grows with the scale factor as, 

\begin{equation}\label{eq:phienergydensityscalefactor}
    \rho_{\phi}(t) \propto a(t)^{-(4+n)}, \quad n> 0,
\end{equation}where $n$ encodes the non-standard cosmological evolution. Notice that for $n=0$, we recover the radiation energy density $\rho_{R}(t) \propto a(t)^{-4}$ that corresponds to the standard case. Then, the scalar field energy density $\rho_{\phi}$ dominates over radiation at early times and redshifts faster than the usual as the Universe expands.

Let us start by assuming that at some period of the early Universe, the Hubble expansion rate was driven by radiation and $\phi$ energy densities. Thus, the total energy density can be written as
\begin{equation}\label{eq:totalenergydensity1}
    \rho = \rho_{R} + \rho_{\phi}. 
\end{equation}
Nevertheless, the radiation energy density $\rho_{R}$ must overcome $\rho_{\phi}$ at a time before BBN to avoid any tension with observations \cite{D_Eramo_2017}. Then, we define the temperature $T_{r}$ at which $\rho_{R}(T_{r}) = \rho_{\phi}(T_{r})$. To be consistent with BBN observations, we impose, 

\begin{equation}
    T_{r} \gtrsim (15.4)^{1/n} \, \mathrm{MeV}.
\end{equation}
When the scalar energy density $\rho_{\phi}$ is negligible, the radiation energy density evolves with temperature as usual with,
\begin{equation}\label{eq:radiationenergydensity}
    \rho_{R}(T) = \frac{\pi^{2}}{30}g_{\star}(T)\,T^{4},
\end{equation}
where $g_{\star}(T)$ is the effective number of relativistic degrees of freedom at temperature $T$.

The next step is to express the energy density of $\phi$ in terms of temperature. Firstly, it is important to remember that the entropy density is given by,
\begin{equation}\label{eq:entropydensity}
        s(T) = \frac{2\pi^{2}}{45}g_{\star\,s}(T)\,T^{3},
\end{equation}where $g_{\star\,s}(T)$ is the effective degrees of freedom of the SM entropy density. Using entropy conservation per comoving volume, $sa^{3} = const.$,  we get, 
\begin{equation}
    a(T) \propto \left(g_{\star}^{1/3}(T)\,T\right)^{-1}.
\end{equation}
Inserting this result into Eq.~\eqref{eq:phienergydensityscalefactor}, and taking the ratio $\rho_{\phi}(T)/\rho_{\phi}(T_{r})$, the $\phi$ energy density can be written as a function of temperature, 
\begin{equation}    
     \rho_{\phi} (T)= \rho_{\phi}(T_{r}) \bigg(\frac{g_{\star\,s}(T)}{g_{\star\,s}(T_{r})}\bigg)^{(4+n)/3}\bigg(\frac{T}{T_{r}}\bigg)^{4+n}.
\end{equation}
Finally, using the definition $\rho_{R}(T_{r}) = \rho_{\phi}(T_{r})$, the total energy density at temperature $T > T_{r}$ becomes,
\begin{align}
    \rho(T) & = \rho_{R}(T) + \rho_{\phi} (T) \nonumber\\ 
    & = \rho_{R}(T)\bigg[ 1 + \frac{g_{\star}(T_{r})}{g_{\star}(T)} \bigg(\frac{g_{\star\,s}(T)}{g_{\star\,s}(T_{r})}\bigg)^{(4+n)/3}\bigg(\frac{T}{T_{r}}\bigg)^{n}\bigg] \label{eq:totalenergydensity2}.
\end{align}
From this equation, we conclude that $\phi$ dominates the Universe for temperatures $T \gtrsim T_{r}$.

Once we know the evolution of the energy density, we can determine the Hubble rate using the Friedmann equation \cite{Kolb:206230},
\begin{equation}
    H = \sqrt{\frac{\rho}{3 M_{Pl}^{2}}},
\end{equation}
where $M_{Pl} = (8\pi G)^{-1/2} = 2.4 \times 10^{18}$ GeV. Assuming that $g_{\star}(T) = g_{\star}$ is a constant for temperatures $T \gg T_{r}$, i.e, for the period in which $\rho_{\phi}$ completely dominates over $\rho_{R}$, the Hubble rate becomes \cite{D_Eramo_2017},
\begin{equation}\label{eq:hubbleratefaster}
    H(T) \approx \frac{\pi}{3}\sqrt{\frac{g_{\star}}{10}}\frac{T^{2}}{M_{Pl}}\bigg( \frac{T}{T_{r}}\bigg)^{n/2}, %\quad T \gg T_{r}.
\end{equation}with $g_{\star} = 106.75$ accounting for the entire SM degrees of freedom. Knowing the Hubble rate in the 
$\Lambda$CDM radiation-dominated cosmology \cite{Barman:2021ifu},

\begin{equation}\label{eq:hubbleradiation}
    H_{R}(T) = \frac{\pi}{3}\sqrt{\frac{g_{\star}}{10}}\frac{T^{2}}{M_{Pl}},
\end{equation}we can rewrite Eq.~\eqref{eq:hubbleratefaster} as,
\begin{equation}
    H(T) \approx H_{R}(T)\bigg( \frac{T}{T_{r}}\bigg)^{n/2},
\end{equation}
where we can explicitly see that for temperatures $T > T_{r}$ non-standard cosmologies ($n>0$) make the Universe expand faster than in the standard radiation-dominated epoch. Setting $T_f$ to be the freeze-out temperature, having a dark matter particle freezing-out in a fast expanding universe, requires $T_{f} > T_{f\,rad}$. Note that from our reasoning for temperatures larger than $T_r$, a non-standard cosmology is at play. 

With the Eq.~\eqref{eq:hubbleratefaster} at hand, we can compute the dark matter relic abundance for the right-handed neutrino $N_{1R}$ when the universe expands too fast. In the standard radiation-dominated scenario, the Boltzmann equation that describes the evolution of the comoving dark matter number density $Y_{N}\equiv n_{N}/s$ is,
\begin{equation}\label{eq:BEQcomovingSTD}
    \frac{dY_{N}}{dx} = -\frac{s\,\langle \sigma v\rangle}{H (x)\,x}(Y_{N}^{2} - Y_{N}^{eq\,2}),
\end{equation}
with $x \equiv m_{N} / T$ and $\langle \sigma v\rangle$ being the thermally averaged annihilation cross-section, and $H(x)$ the standard radiation-dominated Hubble rate. The comoving equilibrium number density can be written as a function of $x$ for a Maxwell-Boltzmann distribution as follows \cite{D_Eramo_2017},
\begin{equation}
    Y^{eq}_{N}(x) = \frac{45\,g_{N}}{4\pi^{4}\,g_{\star\,s}}x^{2}K_{2}(x),
\end{equation}
where $g_{N}$ accounts for the degrees of freedom of the right-handed neutrino dark matter and $K_{2}(x)$ is the modified Bessel function. With $x_r =m_N/T_r$ and $x_f=m_N/T_f$ we obtain,
\begin{equation}
    H(x) \simeq \frac{\pi}{3}\sqrt{\frac{g_{\star}}{10}}\frac{m_{N}^{2}}{M_{Pl}}\,x^{-2}\bigg( \frac{x_{r}}{x}\bigg)^{n/2},
\end{equation}and then from Eq. \eqref{eq:BEQcomovingSTD} we find,

\begin{equation}\label{eq:BEQcomovingfaster}
    \frac{dY_{N}}{dx} = - \frac{A\,\langle \sigma v \rangle}{x^{2-n/2}\,(x^{n} + x^{n}_{r})^{1/2}}(Y_{N}^{2} - Y_{N}^{eq\,2}),
\end{equation}
where $A = \frac{2\sqrt{2}\,\pi}{3\sqrt{5}}\,g_{\star}^{1/2}\,M_{Pl}\,m_{N}$. Approximate analytical solutions are given by \cite{D_Eramo_2017},
\begin{align}\label{eq:comovingnumberdensityfaster}
    Y_{N}(x) & \simeq \frac{x_{r}}{m_{N}\,M_{Pl}\langle \sigma v \rangle}\bigg[ \frac{2}{x_{f}} + \text{log}\bigg(\frac{x}{x_{f}}\bigg)\bigg]^{-1}, \quad n =2 \nonumber \\ 
    Y_{N}(x) & \simeq \frac{x_{r}^{n/2}}{2\,m_{N}\,M_{Pl}\langle \sigma v\rangle} \bigg[ x_{f}^{n/2 - 2} + \frac{x^{n/2-1}}{n-1} \bigg]^{-1}, \quad n > 2. 
\end{align}
\begin{figure}[!ht]
    \centering
    \includegraphics[width=1\linewidth]{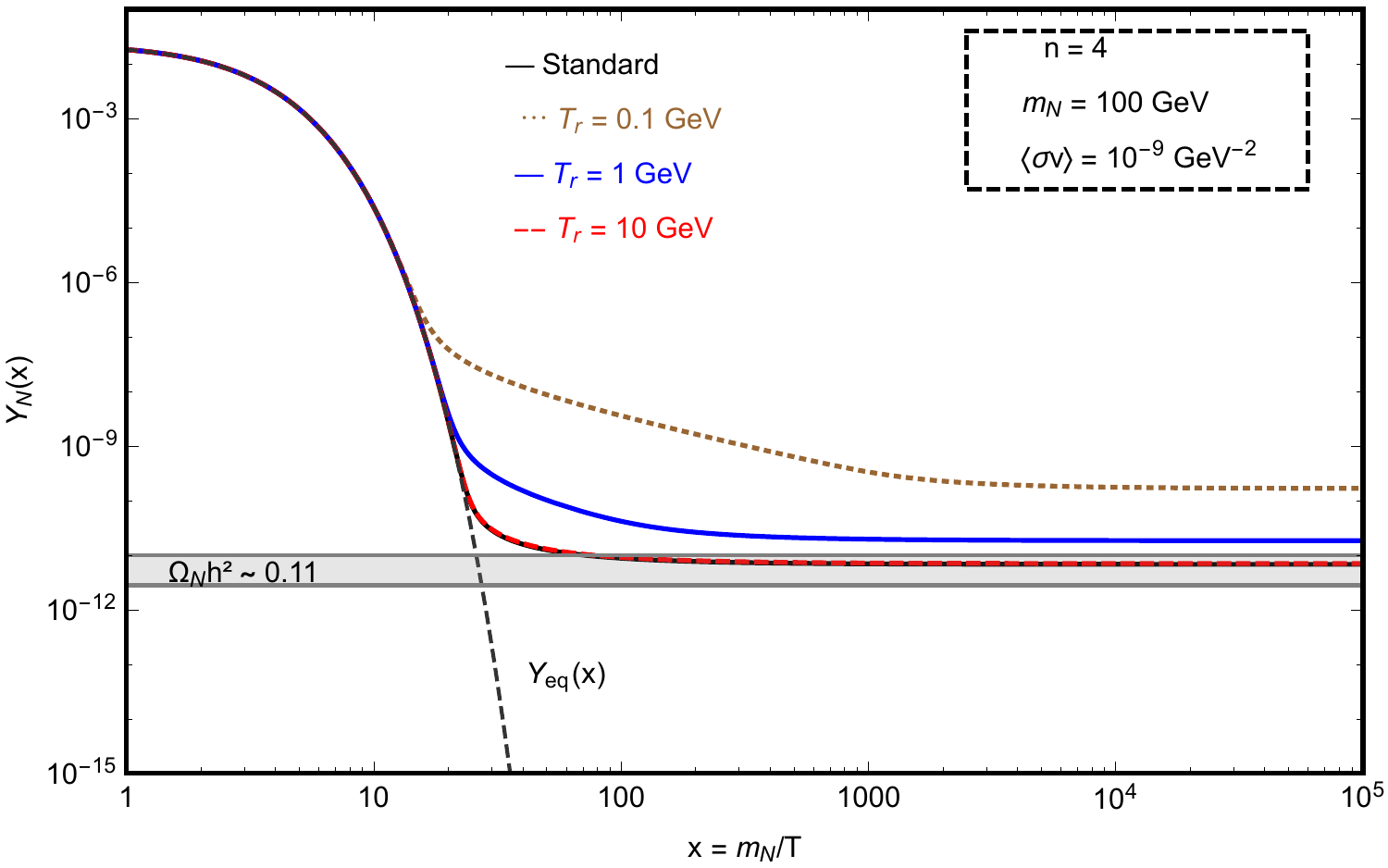}
    \caption{We exhibit of the yield \textit{versus} $x$ for a freeze-out happening during the fast expansion regime. For this case, we choose the dark matter mass $m_N = 100$~GeV and the annihilation cross-section $\langle \sigma v \rangle = 10^{-9}$~GeV$^{-2}$, for the following values of $T_{r} = 0.1$ (brown line), $T_{r} = 1$~GeV (blue line), $T_{r} = 10$~GeV (red line), and $n=4$. For completeness, we also include the yield for the standard case (black line). Note that for $T_{r}=10$~GeV, we have a superposition with the standard case. The shaded gray region is the approximate range for the correct dark matter relic abundance as measured by Planck \cite{2020}. It is clear that the longer the period of $\phi$ dominance, the larger the impact on the yield.}
    \label{fig:yieldfasterdominated}
\end{figure}
Eqs.~\eqref{eq:comovingnumberdensityfaster} are valid for s-wave annihilation processes. In other words, when $\langle \sigma v \rangle$ does not depend on temperature. Thus, these approximate analytical solutions can only be extrapolated up to $x = x_{r}$. In Fig.~\ref{fig:yieldfasterdominated}, we present the numerical result for the yield, assuming the dark matter mass $m_N = 100$~GeV and the annihilation cross-section $\langle \sigma v \rangle = 10^{-9}$~GeV$^{-2}$, for the following values of $T_{r} = 0.1$~GeV (brown line), $T_{r} = 1$~GeV (blue line), $T_{r} = 10$~GeV (red line), and $n=4$. For completeness, we also include the yield for the standard case (black line).

The $n > 0$ cosmologies can produce a higher comoving dark matter number density at the time of freeze-out $x_{f} < x_{r}$. However, for $s$-wave annihilation cross-section and $n \ge 2$, the annihilation rate after freeze-out scales as $\Gamma \propto T^{3}$, whereas $H \propto T^{2 + n/2}$. Hence, the annihilation rate redshifts either slower (or equally) than the Hubble rate. Consequently, after freeze-out the dark matter keeps annihilating until the temperature $T_{r}$, i.e. until radiation starts to govern the universe expansion. In this epoch, $H_{R} \propto T^{2}$, meaning that the dark matter will stop annihilating, leading to a constant energy density. 

\begin{figure}
    \centering
    \includegraphics[width=1\linewidth]{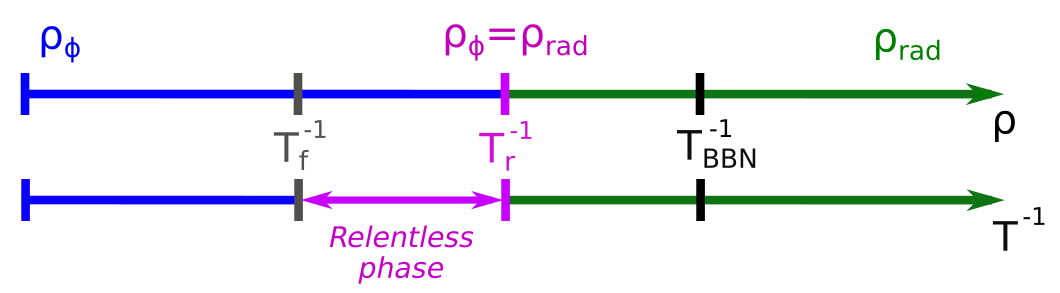}
    \caption{Illustration of the density energy evolution and thermal history for {\it faster early expansion than usual} cosmologies.}
    \label{fig:relentelessdm}
\end{figure}

In Fig. \ref{fig:relentelessdm}, we briefly illustrate the thermal history for the $n \ge 2$ cosmologies. The dark matter decouples at $T_{f}$ whilst the scalar field dominates. The period $T_{r} < T < T_{f}$ defines the {\it relentless} phase in which the dark matter tries relentlessly to go back to thermal equilibrium, unsuccessfully \cite{D_Eramo_2017}. For $n=2$, Eq.~\eqref{eq:comovingnumberdensityfaster} translates the relentless effect via the slow logarithmic decrease of the comoving dark matter number density. Although for $n > 2$, the comoving dark matter number density decreases following a power law. 

Finally, solving Eq. \eqref{eq:BEQcomovingfaster} numerically, we compute the dark matter relic density taking $x \rightarrow \infty$ via the following expression,
\begin{align}\label{eq:DMrelicdensityfaster}
    \Omega_{N}h^{2} & = \frac{s_{0}}{\rho_{0}}\,h^{2}\,m_{N}\,Y_{N}(x \rightarrow \infty) \nonumber \\
    & \simeq 2.82 \times 10^{8}\,m_{N}\,Y_{N}(x \rightarrow \infty),
\end{align}
where $s_{0}$ stands for the SM entropy density today, $\rho_{0}$ the critical energy density, and $h$ is the scale factor for Hubble expansion rate \cite{Zyla:2020zbs}. 

In summary, our dark matter phenomenology is ruled by five free parameters. They encode the interplay between particle physics and cosmology, namely the new gauge coupling, $g_{X}$, the dark matter mass, $m_{N}$, the mediator mass, $m_{Z^{\prime}}$, and the $(n,\,T_{r})$, which are the cosmological input parameters that are related to the energy density of the scalar field that eventually governs the expansion rate of the universe.

We emphasize that the key cosmological input here is the energy density of the scalar field.  It suffices to determine $(n, T_{r})$ and thus describe the entire background in which the dark matter was thermally produced \cite{D_Eramo_2017}. However, many single scalar field cosmologies can reproduce the behavior in Eq. \eqref{eq:totalenergydensity2}. For $n = 2$, with positive scalar potential, it is associated to theories of {\it quintessence} fluids. In our work, we will assume that the energy density scales as $\rho_{\phi} \propto a^{-6}$ in the kination regime \cite{Caldwell_1998, SAHNI_2000, SALATI2003121}. Also, Chaplying gas theories can reproduce the redshift behavior in Eq.~\eqref{eq:totalenergydensity2} and possibly explain dark matter and dark energy \cite{Kamenshchik_2001, Borges:2018evh}.  Although, for $n > 2$, the theories with negative scalar potentials are needed \cite{D_Eramo_2017,PhysRevD.37.3406}.

%%%%%%%%%% Freeze-out During Early Radiation-dominated Era %%%%%%%%%%%%%%%%%

\subsection{\label{ERDtheory}Freeze-out During Early Radiation-dominated Era Followed by a Matter Domination Period}

In a similar vein, we assume the existence of a scalar field $\phi$ that dominates the Hubble expansion rate during some period in the early universe before its decay into SM radiation. However, at this time, $\phi$ behaves as a pressureless fluid\footnote{It redshifts as matter $\rho_{\phi} \propto a^{-3}$.}. After $\phi$ decay, the universe is again dominated by radiation. 

To describe the evolution of this system, we have to use the Boltzmann equations that couples the time evolution of the $\phi$ energy density $\rho_{\phi}$, the SM entropy density $s$ and the dark matter number density $n_{N}$ \cite{Arcadi_2011, Arcadi_2020, Giudice_2001,Drees_2018,Chanda_2020},
\begin{align}
    \frac{d\rho_{\phi}}{dt} & = - 3\,H\rho_{\phi} -\Gamma_{\phi}\,\rho_{\phi}, \label{eq:BEQscalarfieldenergydensityEMD}\\
    \frac{ds}{dt} & = -3\,H\,s + \frac{\Gamma_{\phi}\,\rho_{\phi}}{T} + 2\,\frac{E}{T}\,\langle \sigma v \rangle(n^{2}_{N} - n^{eq\,2}_{N}), \label{eq:BEQSMentropyEMD}\\
    \frac{dn_{N}}{dt} &= - 3\,H\,n_{N} -\langle \sigma v \rangle  (n^{2}_{N} - n^{eq\,2}_{N}), \label{eq:BEQDMnumberdensityEMD}
\end{align}
where $E \simeq m^{2} + 3T^{2}$ is the averaged energy per dark matter particle\footnote{This set of equations does not consider the possibility of the $\phi$ decays also into dark matter particles because we assume $\phi$ does not decay into dark matter. That can be approximately realized, having this scalar field as a singlet with no B-L charges. Anyway, a more general set of equations is provided in \cite{Arias_2019, Arcadi_2020}. Moreover, $\phi$ decaying into dark matter is explored in \cite{Arcadi_2011,Kane_2016, Moroi_2000}.}. From Eqs.~\eqref{eq:BEQSMentropyEMD} and \eqref{eq:BEQDMnumberdensityEMD}, one could see that the injected entropy into SM radiation produced by the decay of $\phi$ can dilute the thermally produced dark matter components. Actually, it happens for any freeze-out scenario before the end of $\phi$ decays \cite{Arias_2019}. 

\begin{figure}[!ht]
    \centering
    \includegraphics[width=1\linewidth]{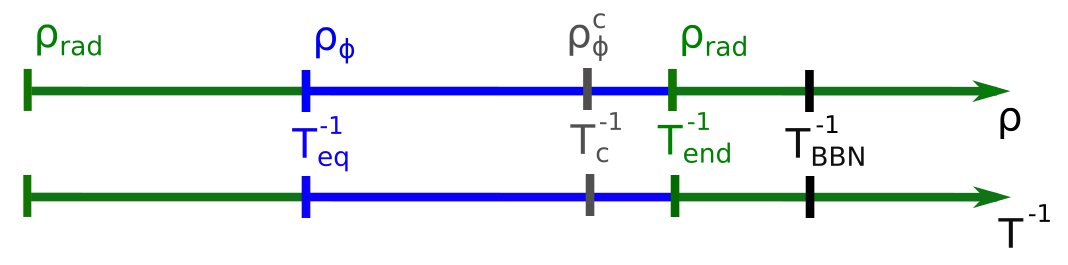}
    \caption{Illustration of the density energy evolution and thermal history for a scalar matter-field dominating in the early Universe. Here, the radiation-dominated dark matter freeze-out, discussed in Section~\ref{ERDtheory}, takes place in the green region (before $T_{eq}^{-1}$), while the matter-dominated freeze-out case, described in Section~\ref{EMDtheory}, happens in the blue region (after $T_{eq}^{-1}$) still before the energy density of the matter field takes a critical value $\rho_{\phi}^c$ at a critical temperature $T_{c}$.}
    \label{fig:matterdominatedillustration}
\end{figure}

In Fig. \ref{fig:matterdominatedillustration}, we briefly recall the early thermal history prior to BBN. The ERD takes place up to the temperature $T_{eq}$ at which the $\rho_{\phi}$ domination starts. Thereafter, $\rho_{\phi}$ effectively dominates over $\rho_{R}$ at the temperature $T_{c}$, and the decay of $\phi$ ends at temperature $T_{end}$ at which the radiation dominates back again. 

The temperature $T_{end}$ is one of the free parameters that specify the cosmological background. It is defined as \cite{Arias_2019, Arcadi_2020}
\begin{equation}
    T_{end} \equiv \bigg[ \frac{90\,M_{Pl}^{2}}{\pi^{2}g_{\star}(T_{end})}\bigg]^{1/4}\,\Gamma_{\phi}^{1/2},
\end{equation}
where $\Gamma_{\phi}$ stands for the total decay rate of $\phi$ into SM radiation. BBN restricts $T_{end} \gtrsim 4\,\text{MeV}$ \cite{2008APh....30..192D, Kawasaki_2000, Hannestad_2004}. The other free parameter is,
\begin{equation}
    \kappa = \frac{\rho_{\phi}}{\rho_{R}}\bigg|_{T = m_{N}},
    \label{eqkappa}
\end{equation}which is the ratio between the energy densities of the scalar field and radiation at $T=m_N$. As we will be varying the dark matter mass, this ratio needs to be recomputed accordingly.

Both parameters $T_{end}$ and $\kappa$ are essentials to determine the dilution resulted from the $\phi$ decay. Notice that the Hubble rate is generally given by,  
\begin{equation}\label{eq:generalHparameter}
    H = \frac{1}{\sqrt{3}\,M_{Pl}}(\rho_{\phi} + \rho_{R} + \rho_{N})^{1/2},
\end{equation}
with the dark matter freeze-out happening at temperature $T_{f}$ during the ERD as usual.

As the dark matter freezes-out at temperature higher than $T_{eq}$, i.e, $T_{eq} \ll T_{f}$, the Hubble in Eq.~\eqref{eq:generalHparameter} is reduced to  Eq.~\eqref{eq:hubbleradiation}, which in terms of the time variable $x$ becomes \cite{Arias_2019},
\begin{equation}\label{eq:hubblerateERD}
    H_{R}(x) \simeq \frac{\pi}{3}\sqrt{\frac{g_{\star}}{10}}\frac{m_{N}^{2}}{M_{Pl}}x^{-2}.
\end{equation}

\begin{figure}[!ht]
    \centering
    \includegraphics[width=1\linewidth]{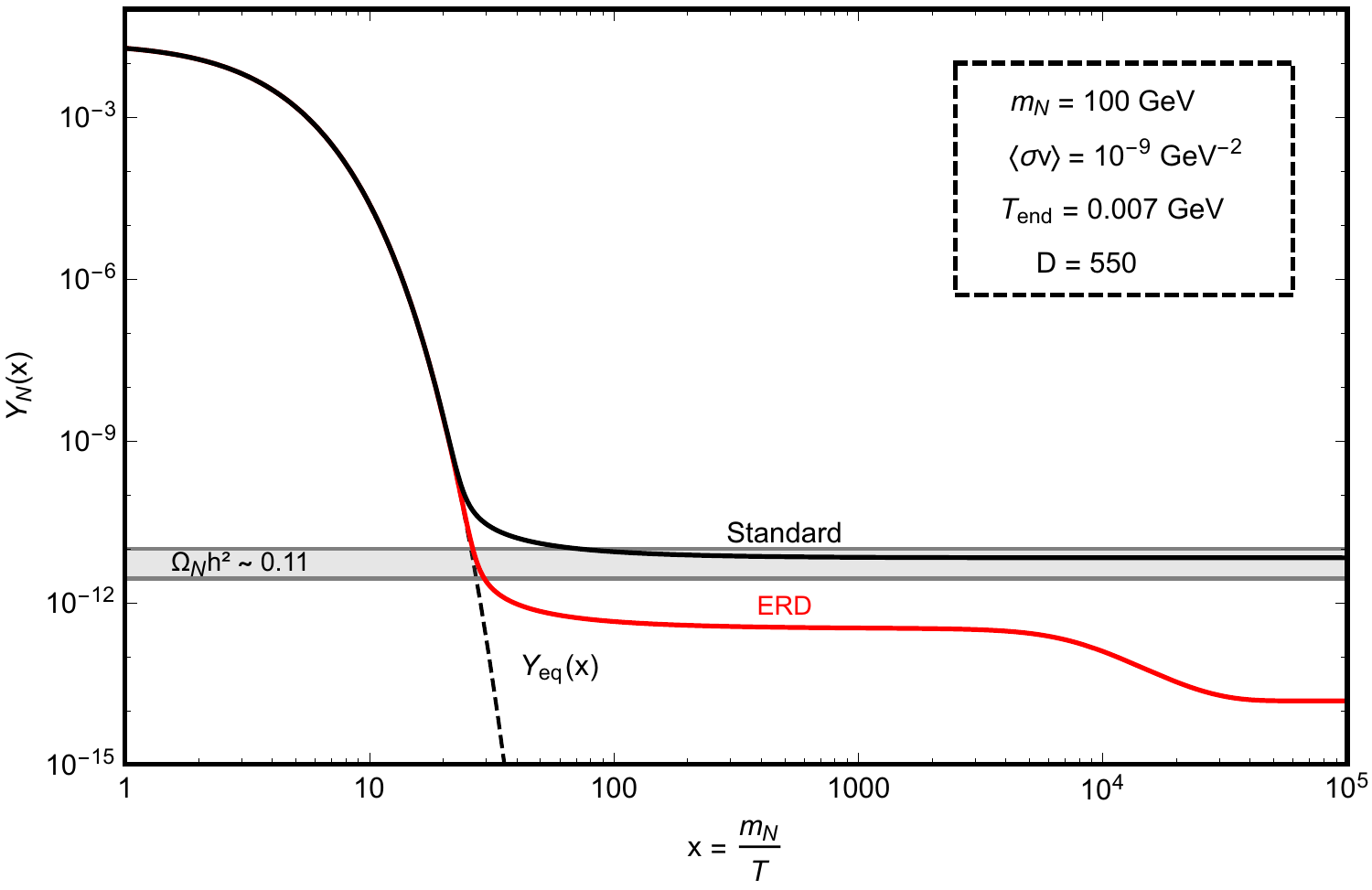}
    \caption{Illustration of the yield \textit{versus} $x$ (red line) provided by a dark matter freeze-out happening during or before a matter-dominated period in the early universe. For this case, we choose the dark matter mass $m_N = 100$~GeV and the annihilation cross-section $\langle \sigma v \rangle = 10^{-9}$~GeV$^{-2}$. Choosing $T_{end} = 0.007$ and $D = 550$. For completeness, we also include the yield for the standard case (black line). Again, the shaded gray region is the approximate range for the correct dark matter relic abundance as measured by Planck \cite{2020}.}
    \label{fig:yeldmatterdominated}
\end{figure}

We emphasize that dark matter decouples much before the decay of $\phi$. Hence, the SM entropy is conserved, $ds/dt = 0$, and Eq.~\eqref{eq:BEQDMnumberdensityEMD} turns into,
\begin{equation}
    \frac{dY_{N}}{dx} = - \frac{s\,\langle \sigma v \rangle}{H_{R}\, x}(Y^{2}_{N} - Y^{eq\,2}_{N}),
\end{equation}which yields,

\begin{equation}\label{eq:comovingDMnumberSTDERD}
    Y_{N}^{std} = \frac{15}{2\pi\sqrt{10\,g_{\star}}}\frac{x_{f}}{m_{N}\,M_{Pl}\,\langle\sigma v\rangle},
\end{equation}
which is the standard solution for the comoving dark matter number density long after the freeze-out and much before $\phi$ decays. In this way, the freeze-out temperature depends neither on $T_{end}$ nor on $\kappa$ \cite{Arias_2019}. As the dark matter abundance is firstly computed in the standard freeze-out case, but it will be changed due to $\phi$ decay (as showed in Fig.~\ref{fig:yeldmatterdominated}), this modification is set by the entropy injection episode that yields a dilution factor defined as,
\begin{equation}
    D \equiv \frac{s(T_{2})}{s(T_{1})} = \bigg(\frac{T_{2}}{T_{1}}\bigg)^{3}.
\end{equation}

The dilution factor is simply the ratio between the SM entropy density at temperatures immediately after $T_{2}$ and before $T_{1}$, the decay of the scalar field. We remind the reader that a similar reasoning is done in textbooks to obtain the temperature difference between photon and neutrinos after $e^+e^-$ annihilations. Assuming the instantaneous decay approximation, the conservation of energy results into,
\begin{equation}
    \rho_{R}(T_{1}) + \rho_{\phi}(T_{1}) = \rho_{R}(T_{2}).
\end{equation}

As $\rho_{\phi}(m) = \kappa \rho_{R}(m)$, and taking $T_{2} = T_{end}$, the dilution factor is,
\begin{equation}\label{eq:dilutionfactor}
    D = \kappa \frac{m_{N}}{T_{end}}.
\end{equation}
In principle, $D$ can take a wide range of values, but  $\kappa$  varies with the dark matter mass, and $T_{end}$ should be larger than $4$~MeV. Moreover, $\kappa$ should be smaller than one at high temperatures to guarantee that the freeze-out happens during a radiation phase. Anyway, combining Eqs.~\eqref{eq:comovingDMnumberSTDERD} and \eqref{eq:dilutionfactor}, we obtain the final comoving dark matter number density \cite{Arias_2019},
\begin{equation}\label{eq:finalDMnumberdensityERD}
    Y_{N} = \frac{Y_{N}^{std}}{D},
\end{equation}and consequently, the overall dark matter relic density is found,
\begin{equation}
    \Omega_{N}h^{2} = \frac{\Omega_{N}^{std}h^{2}}{D},
\end{equation}
where $\Omega_{N}^{std}h^{2}$ stands for the dark matter relic abundance computed in the standard radiation freeze-out scenario. Notice that one cannot randomly pick $D$ at will, pick any dilution factor wished to salvage dark matter models from exclusions, because the value chosen for $D$ is still connected to the dark matter mass and properties of the scalar field. Very large dilution factors are not feasible. Within this production mechanism, we vary the free parameters $(g_{X},\,m_{N},\,m_{Z^{\prime}},\,T_{end},\,D)$ to outline the region of parameter space consistent with existing bounds. In the next section, we obtain the dark matter relic density for freeze-out occurring whilst $\rho_{\phi}$ drives the expansion rate instead.   

%%%%%%%%%% Freeze-out During Early Matter-dominated Era %%%%%%%%%%%%%%%%%%

\subsection{\label{EMDtheory}Freeze-out During Early Matter-dominated Era}
Pre-BBN thermal history is the same as discussed in the last section. Similarly, we consider an unstable scalar field $\phi$ in the early universe which behaves as matter, $\omega_{\phi} = 0$. Moreover, before and after the $\rho_{\phi}$ dominates the expansion, the Universe is radiation-dominated. Now, we investigate the freeze-out process taking place during the period at which the scalar field $\phi$ governs the expansion rate. However, freeze-out happens much earlier than the decay of $\phi$. Taking as reference Fig.~\ref{fig:matterdominatedillustration}, the freeze-out in this case happens between $T_{eq}^{-1}$ and $T_c^{-1}$.

The Hubble parameter is found in Eq.~\eqref{eq:generalHparameter}. Although, we define a temperature $T_{\star}$ at which the $\rho_{\phi}$ begins to evolve. It allows us to parameterize the relative fraction of the total energy density at $T = T_{\star}$ in terms of \cite{Hamdan_2018},
\begin{equation}
    r \equiv \frac{\rho_{R} + \rho_{N}}{\rho_{\phi} +\rho_{R} + \rho_{N}} \bigg|_{T=T_{\star}} = \bigg[ 1+ \frac{g_{\phi}(T_{\star})}{g_{\star}(T_{\star}) + g_{N}}\bigg( \frac{m_{\phi}}{T_{\star}}\bigg)^{4}\bigg]^{-1}
    \label{eqr}
\end{equation}
where $r \in [0,1]$. In this way, for $T > T_{\star} > m_{N}$, we have,
\begin{equation}\label{eq:hubbleratioEMD}
    \frac{H^{2}}{H^{2}_{\star}} = \frac{g_{\star}\, r}{g_{\star}+g_{N}}\bigg( \frac{a_{\star}}{a}\bigg)^{4} + (1-r)\bigg( \frac{a_{\star}}{a}\bigg)^{3}+\frac{g_{N}\,r}{g_{\star}+g_{N}}\bigg( \frac{a_{\star}}{a}\bigg)^{4},
\end{equation}
where $H_{\star} \equiv H(T_{\star})$ and $a_{\star} \equiv a(T_{\star})$. The conservation of entropy leads to,
\begin{equation}
    \frac{a_{\star}}{a} \simeq \bigg(\frac{g_{\star}(T)}{g_{\star}(T_{\star})}\bigg)^{1/4}\bigg( \frac{T}{T_{\star}}\bigg).
\end{equation}
Inserting this result into Eq.~\eqref{eq:hubbleratioEMD}, the Hubble parameter becomes approximately,
\begin{equation}\label{eq:HubbleEMDcase}
    H \simeq H_{\star}\bigg(\frac{g_{\star}(T)}{g_{\star}(T_{\star})}\bigg)^{3/8}\bigg( \frac{T}{T_{\star}}\bigg)^{3/2}\bigg[ (1-r) + r\bigg(\frac{T}{T_{\star}}\bigg) \bigg]^{1/2}.
\end{equation}

Thus, a matter-domination phase arises immediately at $T_{\star}$, (See Eq.~\eqref{eqr}) for $r \ll 1$, which leads to $H \propto T^{3/2}$. For $r \simeq 1$, the Hubble rate goes as $H \propto T^{2}$, which is associated to a radiation-domination epoch. Anyway, the Universe is radiation-dominated at $T_{\star}$, and the matter-domination phase will naturally arise at temperature,
\begin{equation}
    T_{eq} \equiv T_{\star}\,\bigg( \frac{a_{\star}}{a(T_{eq})}\bigg) = \frac{1-r}{r} \bigg[\frac{g_{\star}(T_{\star})}{g_{\star}(T_{eq})} \bigg]^{1/4}\,T_{\star},
\end{equation}
at which $\rho_{\phi}$ accounts for $50\%$ of the total density energy in agreement with \cite{Hamdan_2018,Arias_2019}.

The matter field that drives the expansion decays only when it effectively dominates over radiation at temperature,
\begin{equation}\label{eq:phidecaytemperature}
    T_{c} \simeq (1-r)^{-1/3}\,\bigg[\frac{g_{\star}(T_{end})}{g_{\star}(T_{\star})}\,\frac{T_{end}^{4}}{T_{\star}} \bigg]^{1/3}.
\end{equation}
We stress that the freeze-out takes place in a matter-domination period, which occurs much before the scalar field decays. In other words, the freeze-out temperature should lie in the range $T_{c} \ll T_{f} \ll T_{eq}$ \cite{Arias_2019,Hamdan_2018}.% recall $T_{\Gamma} = T_{c}$ and $T_{end}$ = T_{RH}.

We solve the Boltzmann equation for a matter-dominated universe, Eq.~\eqref{eq:BEQDMnumberdensityEMD}, with the Hubble given in  Eq.~\eqref{eq:HubbleEMDcase}. Taking the limiting case of matter-domination, $r \ll 1$, and assuming a $s$-wave annihilation cross-section, we find the comoving dark matter number density after freeze-out long after the matter-domination epoch \cite{Hamdan_2018,Hamdan:2018fqj},
\begin{equation}
    Y_{N}^{MD} = \frac{3}{2}\sqrt{\frac{45}{\pi}}\frac{\sqrt{g_{\star}}}{g_{\star\,s}}\frac{x_{f}^{3/2}}{m_{N}\,M_{Pl}\,\langle\sigma v\rangle\,x_{\star}^{1/2}},
\end{equation}
where $x_{\star} = m_{N}/T_{\star}$ and $g_{\star\,s} \simeq g_{\star} \simeq cte$ at the time of freeze-out. However, it will be diluted due to entropy injection of the scalar field decay. In a similar vein, the dilution factor is the ratio between the entropy density computed at temperatures immediately before and after  the scalar field decay, which in terms of $T_{end}$ and $T_{\star}$ leads to, \footnote{Notice that $\zeta$ may be equal to $D^{-1}$, but for different freeze-out scenario and different parameters.} \cite{Hamdan:2018fqj},
\begin{equation}
    \zeta = \frac{s(T_{1})}{s(T_{2})} \simeq (1-r)^{-1}\,\frac{g_{\star}(T_{c})}{g_{\star}(T_{\star})}\,\frac{T_{end}}{T_{\star}}.
\end{equation}
Therefore, the comoving dark matter number density becomes,
\begin{equation}
    Y_{N} = \zeta\, Y_{N}^{MD},
\end{equation}and consequently,
\begin{equation}
    \Omega_{N}h^{2} = \zeta\,\Omega_{N}^{MD}h^{2},
\end{equation}which is the dark matter relic density for a particle that freezes-out during a matter-domination era, which later experiences the decay of the scalar field $\phi$. This scenario can also be represented by Fig.~\ref{fig:yeldmatterdominated}, where after freeze-out we see a dilution in the dark matter yield. In the next section, we present the constraints for our particle physics model.

%%%%%%%%%%%%%%%%%%%%%%%%%%%%%%%%%%%%%%%%%%%%%%%%%%%%%%%%%%%%%%%%%%%%%%%%%%
%%%%%%%%%%%%%%%%%%%%% Constraints and Results %%%%%%%%%%%%%%%%%%%%%%%%%%%%
%%%%%%%%%%%%%%%%%%%%%%%%%%%%%%%%%%%%%%%%%%%%%%%%%%%%%%%%%%%%%%%%%%%%%%%%%%
\section{Constraints}

As the model is based on a 2HDM fashion there are several constraints which have been derived specifically to the canonical 2HDM, but some are also applicable to our model with proper adjustments. We will cover one by one below.

\subsection{Collider}

The most relevant collider limits stem from $Z^\prime$ searches at the LHC. As the $Z^\prime$ coupling to SM fermions is not suppressed, such $Z^\prime$ boson can be easily produced at the LHC, giving rise to either dijet or dilepton signal events. If the $Z^\prime$ coupling to dark matter were different and larger than the one with SM fermions, the invisible decay of the $Z^\prime$ boson could be significant to weaken the LHC lower mass bound. However, in our model, the $Z^\prime$ field interacts with equal strength to all fermions. Thus, the dilepton and dijet searches at the LHC are not meaningfully impacted by the presence of $Z^\prime$ decays into dark matter pairs \cite{ATLAS:2019erb,CMS:2019emo}. 
These two datasets have been considered, and the respective bound are displayed in our plots.

\subsection{Flavor Physics}

Interestingly, our model can also be constrained by flavor physics observed despite the absence of flavor changing interactions. The charged Higgs boson contributes to the $b \rightarrow s \gamma$ at one-loop level via the Cabibbo–Kobayashi–Maskawa matrix \cite{Grinstein:1990tj,Misiak:2017bgg}. The limit is reported in terms of $\tan\beta$ and the charged scalar mass. In our model, the mass of the charged Higgs is proportional to the $v_s$, vev of the singlet scalar, and $\tan\beta$. Taking $\tan\beta=1$, we can convert the lower mass limit on the charged Higgs mass into a bound on $v_s$. Knowing that the $Z^\prime$ mass is controlled by $v_s$ and $g_{X}$, for a given value of $g_{X}$ we now have a constraint on the $Z^\prime$ mass. We exhibit this limit in our plots. 

\subsection{Atomic Parity Violation}

Atomic parity violation (APV) effects are often overlooked in physics beyond the Standard Model endeavours \cite{Davoudiasl:2012qa,Davoudiasl:2012ag,Davoudiasl:2014kua,Arcadi:2019uif}. The $Z^\prime$ gauge boson rising from the B-L gauge group has only vectorial interactions with fermions. In principle, that would lead to no APV, but in the presence of mass mixing with the Z boson this is no longer true. Atomic transitions in Cesium has been proved to be a great laboratory to probe $Z^\prime$ contributions to APV via mass mixing. Following \cite{Arcadi:2020aot}, the contribution to the weak charge of Cesium in our model is found, 

\begin{equation}
\Delta Q_W =-59.84 \delta^2 - 220\delta \sin\theta_W \cos\theta_W \epsilon \frac{m_Z}{ m_{Z^\prime}}-133\delta^2\tan\beta^2    
\end{equation}where $\delta$ is the mass mixing parameter,
\begin{equation}
    \delta=\frac{\cos\beta \cos\beta_d}{\sqrt{1-\cos^2\beta \cos^2\beta_d}}
\end{equation}where we have defined $\tan\beta_d=\frac{v_1}{v_s}$ \cite{Campos:2017dgc}. As the $v_s$ controls the $Z^\prime$ mass, with $\tan\beta=1$, we can display this limit in terms of the $Z^\prime$ mass for a given $g_{X}$. We show this constraint in our plots.

\subsection{Direct Detection}

Dark matter particles might scatter off of nuclei, leaving a signal at direct detection experiments. We are considering heavy mediators only, thus the momentum transfer in the scattering process is much smaller than the mediator mass. Therefore, the dark matter interaction with quarks can be treated using effective operators. Our model features the effective Lagrangian,
\begin{equation}
\label{eqDD_SI}
    \mathcal{L}_{DD}=\frac{g_X^4}{18 m_{Z^{'}}^2}\bar N_{1R} \gamma^\mu \gamma_5 N_{1R} \bar q \gamma_\mu q.
\end{equation}where $q=u,d$. We point out that only the valence quarks contribute due to the vector current present in Eq.~\eqref{eqDD_SI}. 

In models where $Z-Z^\prime$ mass mixing is absent, Eq.~\eqref{eqDD_SI} represent the only relevant source of direct detection signal. Interestingly, such term does not give rise to the standard spin-independent (SI) or spin-dependent (SD) dark matter-nucleon scattering signal. Hence, we need to map this Lagrangian onto a more general formalism using effective operators  \cite{Fitzpatrick:2012ix,Fitzpatrick:2012ib,Anand:2013yka} to understand that this interaction gives rise to the operators,

\begin{equation}
    \bar N_{1R} \gamma^\mu \gamma_5 N_{1R} \bar q \gamma_\mu q \rightarrow 2 \vec{v}_N^{\bot} \cdot \vec{S}_N+2 i \vec{S}_N \cdot \left(\vec{S}_n \times \frac{\vec{q}}{m_N}\right)
    \label{opSI}
\end{equation}where $\vec{S}_{N}=1/2$ is the right-handed neutrino spin, $\vec{S}_{n}$ the net nucleus spin, $\vec{v}_N^\bot=\vec{v}+\frac{\vec{q}}{2 \mu}$, $\vec{q}$ is the momentum transfer, and $\mu$ the reduced mass.

The first term in Eq.~\eqref{opSI} features an enhancement with the atomic mass but is velocity suppressed, whereas the second is momentum suppressed and spin-dependent. That said, the first term will be dominant but will yield a distinct energy spectrum at the detector. We properly account for behavior using the code described in \cite{Cirelli:2013ufw}.

However, our model has a second Higgs doublet charged under the new gauge symmetry. Therefore, the $Z-Z^\prime$ mass mixing in unavoidable. This mixing induces the operator,

\begin{equation}
\label{eqDD_SD}
    \mathcal{L}_{DD} = \frac{1}{\Lambda^2} \bar N_{1R} \gamma^\mu \gamma_5 N_{1R} \bar q \gamma_\mu \gamma^5 q.
\end{equation}where $\Lambda$ here encodes both the Z mediator and $Z^\prime$ mediator via t-channel diagrams. This axial-vector interactions is known to generate a spin-dependent scattering which is found to be,
\begin{widetext}

\begin{equation}
    \sigma_{\rm N p}^{\rm SD}=\frac{\mu^2_{N p}}{4\pi}\left \vert \left(\frac{g^A_{Zu}g_{ZN}}{m_Z^2}+\frac{g^A_{Z^\prime u}g_{Z^\prime N}}{m_{Z^\prime}^2}\right)\Delta_u^p+\left(\frac{g^A_{Zd}g_{ZN}}{m_Z^2}+\frac{g^A_{Z^\prime d}g_{Z^\prime N}}{m_{Z^\prime}^2}\right) (\Delta_d^p+\Delta_s^p) \right \vert^2
\end{equation}where $g^A_{Z\,u,d}$ ($g^A_{Z^\prime\,u,d})$ are the axial couplings of the $Z$ ($Z^\prime$) boson with quarks, while $g_{ZN}$ ($g_{Z^\prime N}$) are the axial couplings of the $Z$ $(Z^\prime)$ boson with dark matter. The explicit expressions can be extracted directly from Eq.~\eqref{zzgeralcoma1}, and $\Delta_{u,d,s}^p$ are form factors accounting for the light quarks contributions to the nucleon spin \cite{Fitzpatrick:2012ix,Fitzpatrick:2012ib}.

\end{widetext}

\section{\label{constrainsandresults}Results}
\begin{widetext}

\begin{figure}[!ht]
    \centering
    \includegraphics[width=\textwidth]{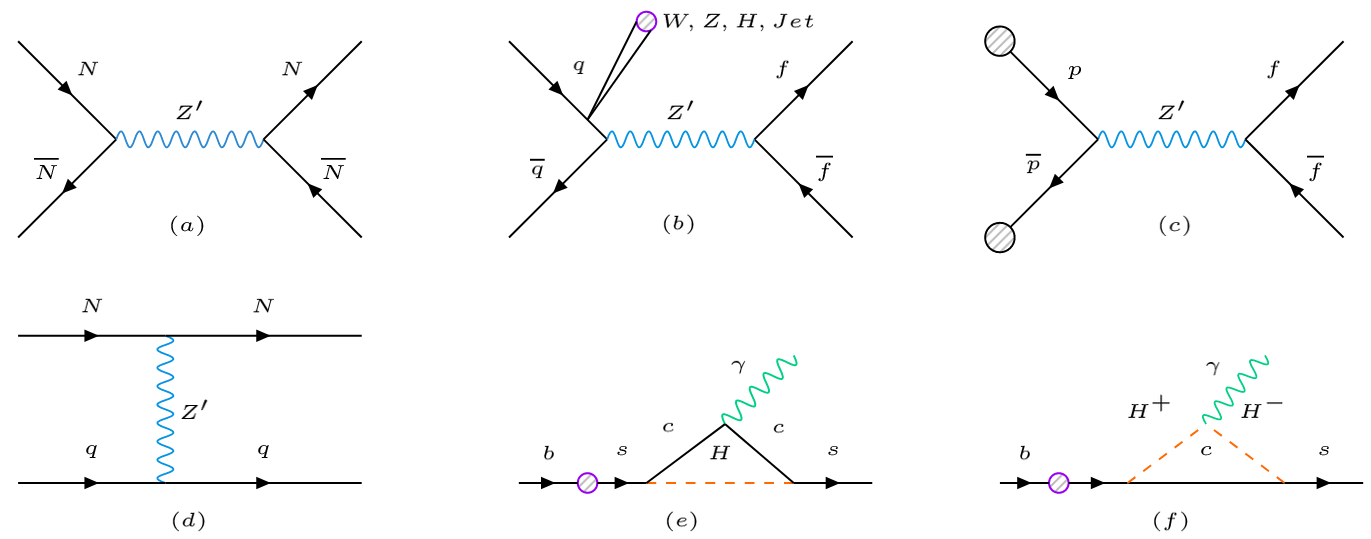}
    \caption{Here we have the Feynman diagrams related to the dark matter phenomenology of this work. The diagram (a) gives the dark matter relic abundance, (b) corresponds to collider searches, (c) is associated to the atomic parity violation bounds, (d) to direct detection searches, and (e) \& (f) comes from flavor physics.}
    \label{fig:feyn-diag}
\end{figure}

\end{widetext}

In this section, we present the results for each scenario described above. We compute the dark matter relic density for different cosmological setups and compare aforementioned bounds. The results are presented in the $m_{Z^{\prime}}$ vs $m_{N}$ plane. We cover several cosmological scenarios, varying their cosmological parameters.  We also fixed the gauge coupling, $g_{X} = 1$, since the main focus here is to study the impact of different cosmological parameters on this model. In any case, decreasing the gauge coupling $g_X$ will provide lower cross-sections, leading to higher relic densities. To compensate for this change and to get the right relic density, we have to diminish the $Z^\prime$ mass, usually leading to more constrained scenarios\footnote{For a detailed study on the impact of the gauge coupling on the results we recommend \cite{Arcadi:2020aot}.}. For completeness, we also include the main Feynman diagrams responsible for the dark matter relic density, collider and direct searches, atomic parity violation and flavor physics in Fig.~\ref{fig:feyn-diag}.

We assess the impact of non-standard cosmology by comparing our findings with the one stemming from standard freeze-out. The colorful shaded regions represent the bounds from direct detection (blue region), dijet (red region), dilepton (cyan region), flavor physics (light green), APV (dark green region), and perturbative unitarity (magenta). We overlay the curves that delimit the parameter, yielding the correct relic density within the standard freeze-out (solid black).

\begin{figure*}[!ht]
	\includegraphics[width=0.48\linewidth]{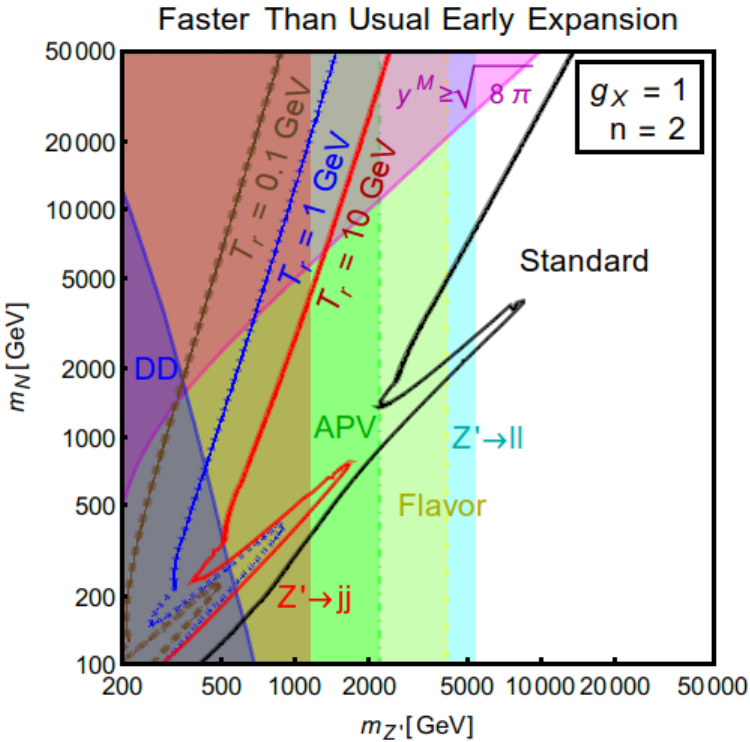}
	\includegraphics[width=0.48\linewidth]{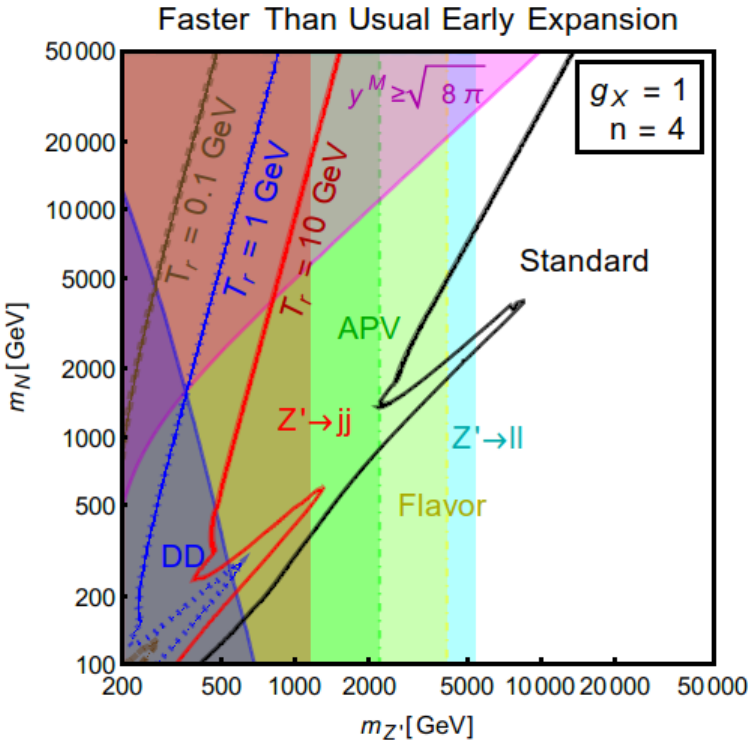}
	\includegraphics[width=0.48\linewidth]{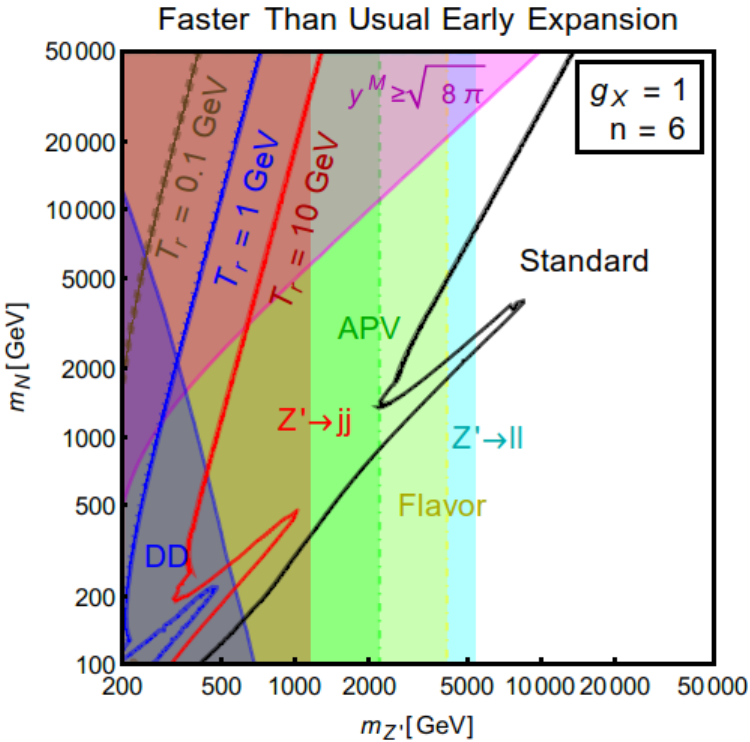}
	\caption{Fast Expansion Case: We exhibit the curves that delimit the parameter space that leads to the correct relic density for $T_r=0.1$ (brown), $1$~GeV (blue) and $10$~GeV (red). We also display the region which reproduces the correct relic density within the standard freeze-out for comparison. The colorful shaded regions represent the bounds from direct detection (blue region), dijet (red region), dilepton (cyan region), flavor physics (light green), APV (dark green region), and perturbative unitarity (magenta).}
	\label{fig:plotsfaster}
\end{figure*}

\subsection{\label{resultfaster}Faster than usual early expansion}

In the standard freeze-out case, the right-handed neutrino dark matter can nicely reproduce the correct relic density and evade existing bounds around the $Z^\prime$ resonance, and away from the resonance when $m_N > m_{Z^\prime}$. In the context of a faster expansion rate at early times, the $(n,\,T_{r})$ parameters have an enormous impact on the dark matter relic abundance \cite{D_Eramo_2017,barman2021scalar}. Moreover, the dark matter mass also plays a key role. We explore $n=2, 4, 6$ cosmologies, for temperatures $T_{r} = 0.1,1,10$ GeV to have an overall idea of how a faster expansion affects the standard relic density. The largest temperature at the end of the relentless phase should be $T_{r}=10$~GeV, because we must enforce $T_{f} > T_{r}$ to keep freeze-out occurring during the scalar field domination. That said, we exhibit in Fig. \ref{fig:plotsfaster} the region of parameter in the $m_N$ vs $m_{Z^\prime}$ plane that yields the correct relic density. For comparison, we also display with a black curve the region of parameter space that gives the right relic density in the standard freeze-out regime. For all plots we took $g_{X}=1$, in the top-left panel we adopted  $n=2$, in the top-right panel we assumed and $n=4$, and in the bottom plot and $n=6$.

It is clear that the parameters $n$ and $T_{r}$ can enhance the dark matter relic density.  We explored an interplay between $n$ and $T_r$, keeping one fixed and varying the other. In this cosmological background, a faster expansion, which enhances the dark matter abundance, the parameter space that reproduces the correct relic density is shifted to smaller $Z^\prime$ masses. One can understand this conclusion, remembering that the $Z^\prime$ mass enters the denominator of the thermal annihilation cross-section. Hence, the smaller the $Z^\prime$ mass, the smaller the abundance. This shift to lower $Z^\prime$ masses appears to counterbalance the fast expansion present during the dark matter freeze-out. As we increase further the expansion rate and take $n=4$ (top-right-panel), and $n=6$ (bottom-panel), the regions that reproduce the correct relic are also further shifted to smaller $Z^\prime$ masses. Furthermore, for each cosmological background $n$, the larger the ratio $x_{r} = m_{N}/T_{r}$, the longer the relentless phase. Then, for a fixed $T_{r}$, the heavier right-handed neutrino dark matter particles, the larger $Z^{\prime}$ masses (smaller cross-sections) to compensate the suppression to the dark matter relic density due to their longer time relentlessly annihilating.

The slope of the curves is mostly governed by $n$ as it is closely related to the equation of state of the field $\phi$ \cite{D_Eramo_2017}. Notice that this fast expansion history is completely excluded by existing bounds, even in the $Z^\prime$ resonance. In the face of this, we will explore the impact of the early matter dominated field on this model. 

%notice that the channels that favor $m_{N} > m_{Z^{\prime}}$ regime becomes relevant after the resonances: $ZH$, $W^{+}H^{-}$, $H_{s}H_{s}$ and $Z^{\prime}Z^{\prime}$.

\subsection{Matter-Field Domination in the Early Universe}

In this section, we deal with the non-standard freeze-out scenarios discussed in Sections \ref{ERDtheory} and \ref{EMDtheory}. Here, we use the cosmological parameters $(T_{end},\, D)$ and $(T_{\star},\,\zeta)$ for ERD and EMD, respectively. We firstly analyze the early radiation-dominated freeze-out, thereafter, the dark matter decoupling during EMD in which the field $\phi$ drives the Hubble expansion, but it is still away from its decay.

\paragraph{Freeze-out During Early Radiation-dominated Era Followed by a Matter Domination Period}
In this case, we explore $D = 550$ and $2750$, fixing $T_{end} = 0.007$ GeV \cite{Arcadi_2020}. As discussed earlier, $\kappa$ is not fixed in order to adjust to the dark matter mass scales (See Eq.\eqref{eq:dilutionfactor}). It is common to evoke a constant and ad hoc dilution factor to bring down the dark matter relic density and circumvent existing bounds. Notice that this dilution factor cannot take any value, because it does depend on $T_{end}$ and $\kappa$. Naively, one might think that $\kappa$ and $T_{end}$ are completely independent parameters. As $\kappa$ is the ratio between the $\phi$ and radiation densities, thus depends on $g_{\ast}(T_{end})/g_{\ast} (T=m_{N})$ \cite{Arias_2019}. Therefore, $\kappa$ does feature a dependence on $T_{end}$. As we need to impose $T_{end}>4$~MeV due to BBN bounds, clearly the dilution factor $D$ cannot take arbitrarily large values.

\begin{figure}[!ht]
	\centering
	\includegraphics[width=1.\linewidth]{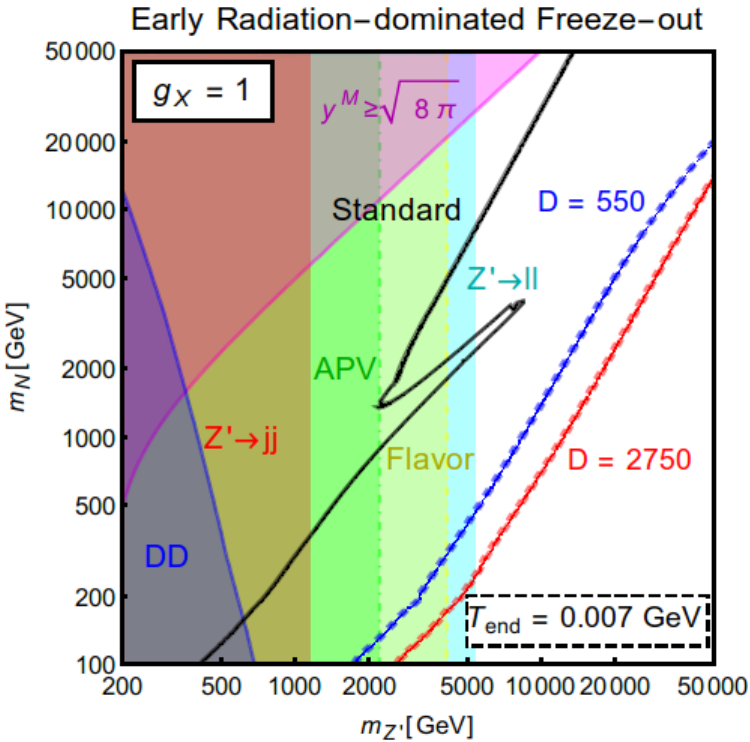}
	\caption{The correct dark matter relic abundance for dark matter freeze-out happening during the early radiation-dominated epoch. For $g_{X} = 1$, the cosmological parameters are fixed to $T_{end} = 0.007$~GeV and $D = 550$ and $2750$ (blue and red curves, respectively). The constraints are the same as in Fig.\ref{fig:plotsfaster}.}
	\label{fig:resultradiation}
\end{figure}

In Fig. \ref{fig:resultradiation}, we see that the contours that delimit the region of parameter space which yield the right dark matter abundance are free from existing bounds even away from the resonance, for $m_{N} < m_{Z^{\prime}}$. Notice that the larger the entropy after the decay of the scalar field quantified by the dilution factors $D=550\,\mathrm{and}\, 2750$, the smaller the cross-sections and the lighter the dark matter masses are required to obtain the correct relic density. In other words, there is a shift towards heavier $Z^{\prime}$ bosons and smaller dark matter masses. It is important to emphasize that we are fixing the value of $T_{end}=0.007$ and leaving the $\kappa$ parameter free to get the right value of the dilution factor. If we fix the mass $m$ and the $\kappa$ parameter, $T_{end}$ will be inversely proportional to the dilution factor $D$, according to the Eq.~\eqref{eq:dilutionfactor}. Conversely, the region of parameter space that reproduces the correct relic density in the standard freeze-out is rather more constrained. Hence, a successful way to salvage WIMP models from restrictive direct detection and collider bounds is to assume a standard freeze-out followed by a short matter domination stage governed by the scalar $\phi$, which then decays and injects entropy, shifting the relic density curve to the right side in Fig. \ref{fig:resultradiation}.

\begin{figure}[!ht]
	\centering
	\includegraphics[width=1\linewidth]{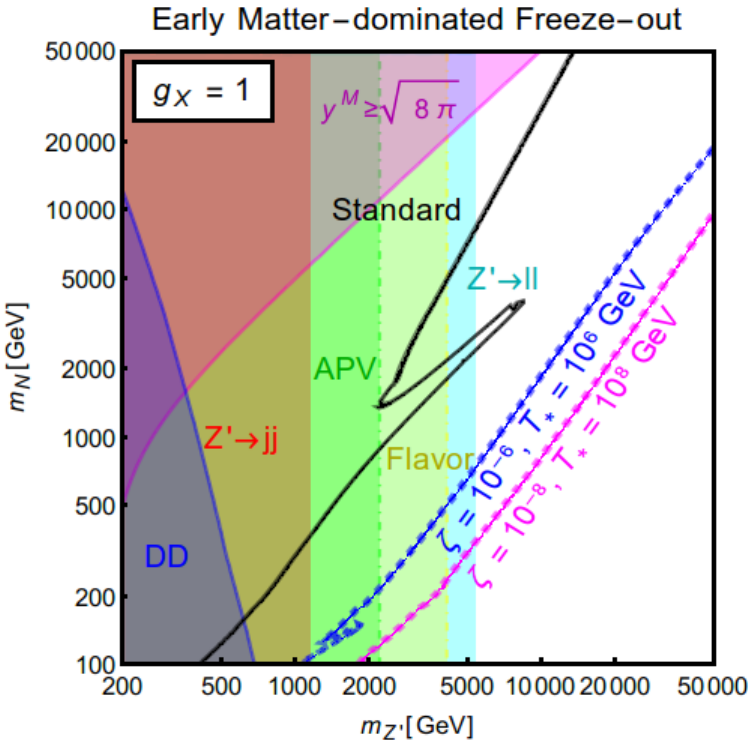}
	\caption {The correct dark matter relic abundance for dark matter freeze-out happening during the matter-dominated epoch before the decay of $\phi$. For $g_{X} = 1$, the cosmological parameters are set into pairs of $\zeta\,\mathrm{and}\, T_{\star}$. We have chosen $\zeta = 10^{-6}\,\mathrm{and}\,T_{\star} = 10^{6}$~GeV (blue curve), and $\zeta = 10^{-8}\,\mathrm{and}\,T_{\star} = 10^{8}$~GeV (magenta curve). The constraints are the same as in Fig.~\ref{fig:plotsfaster}.}
	\label{fig:resultmatter}
\end{figure}

\paragraph{Freeze-out During Early Matter-dominated Era}
In Fig.~\ref{fig:resultmatter}, we show the results for the early matter-dominated freeze-out. We have chosen $\zeta = 10^{-6}\,\mathrm{and}\,T_{\star} = 10^{6}$~GeV, and $\zeta = 10^{-8}\,\,\mathrm{and}\,\,T_{\star} = 10^{8}$~GeV. These choices are in agreement with the freeze-out during EMD condition, namely, $T_{c} \ll T_{f} \ll T_{eq}$. We took $T_{end} \simeq 10$~GeV, and $r = 0.99$. The blue and magenta curves delimit the region of parameter space that yields the correct relic density. It is visible that most region of parameter space is now consistent with existing bounds.

Our results comparing with the standard case (black lines), leads to scenarios completely free from the bounds, for dark matter masses larger than $200$~GeV, for both ERD and EMD cases. On the other side, for the standard case, the only regions that survive the limits are near the $Z^\prime$ resonance. Therefore, for our model, the search for dark matter masses around the $TeV$-scale may hint the presence of non-standard cosmological histories in the early universe.

\subsection{The relentless phase {\it versus} the entropy injection}
\begin{figure}[!ht]
	\centering
	\includegraphics[width=1\linewidth]{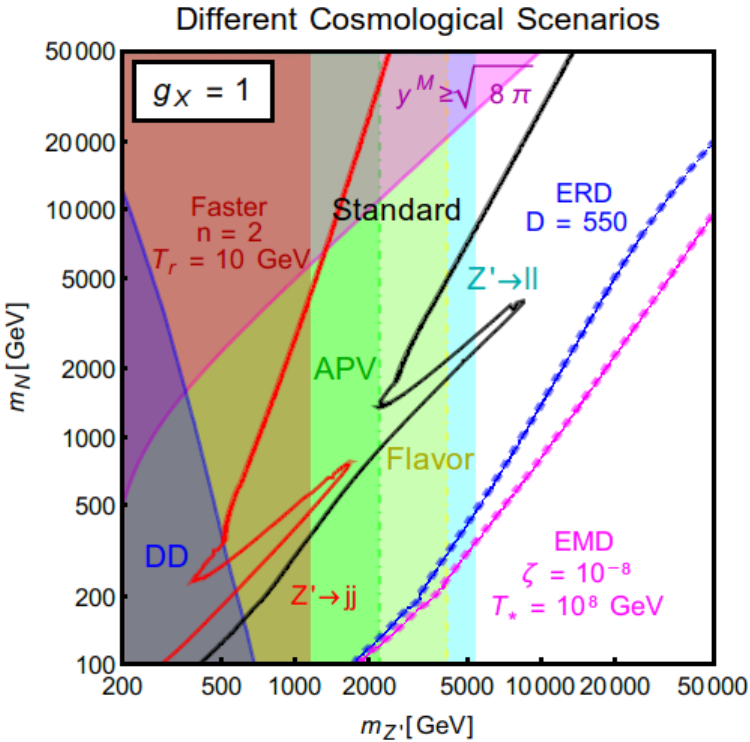}
	\caption{The correct dark matter relic abundance is presented for the {\it quintessence} fluid $n=2$ (solid red contour), the early radiation-dominated freeze-out case (blue contour) and for the freeze-out during the EMD (magenta contour). We took $T_{r} = 10$~GeV, $D=550$, and $\zeta=10^{-8}$ and $T_{\star} = 10^8$~GeV, respectively. The constraints are still the same as in Fig.\ref{fig:plotsfaster}.}
	\label{fig:resultall}
\end{figure}

In this section, we put all results into perspective and discuss the key findings concerning the overall dark matter density in the presence of non-standard cosmologies. In Fig.~\ref{fig:resultall}, we present the results for the scenarios in which a quintessence fluid (red curve) with $n=2$ and $T_{r} = 10$~GeV dominates the expansion rate (Faster Expansion), the standard early radiation domination occurs (ERD), but it is followed by the decay of a matter field for $D = 550$ with $T_{end} = 0.007$~GeV, and lastly the freeze-out during early matter domination (EMD) for $\zeta = 10^{-8}$ and $T_{\star} = 10^{8}$~GeV. 

It is remarkable that for a quintessence-dominated freeze-out, the contour arises in the left-hand side of the standard freeze-out. This reflects the fact that the larger Hubble rate, the larger the annihilation cross-section to reproduce the correct relic density.

Instead, for freeze-out happening during periods in which the Hubble rate is equivalent to or slower than radiation, the contour appear in the right-hand side, which is less constrained by data. Hence, the presence of a scalar field that drive the Hubble rate during or after the dark matter freeze-out represents an important direct to be explored by the next generation of experiments.

\section{\label{dc} conclusions}

In this work, we investigated the dark matter phenomenology of a weak scale right-handed neutrino dark matter under three different cosmological settings. We concretely incorporated this right-handed neutrino into a Two Higgs Doublet Model featuring an Abelian gauge symmetry. The model itself is well-motivated for being able to account for neutrino masses and absence of flavor changing interactions in the scalar sector.

The dark matter phenomenology is driven by the presence of a $Z^\prime$ field, as well as by the $Z-Z^\prime$ mass mixing that arises because a Higgs doublet is involved in the spontaneous symmetry breaking of the new Abelian gauge symmetry. Concerning non-standard cosmology, we explored the case where the universe expands faster than usual, and the scenarios where the dark matter freeze-out takes place during a matter domination epoch, as well as during a usual radiation domination, but it is then followed by a matter domination phase. 

For the faster expanding case, we found a very constrained scenario, since larger cross-sections are necessary to reproduce the right relic density, whereas when a matter field dominates the expansion rate in the early universe either during or after the dark matter freeze-out the relic density curves are shift away from existing bounds.  
After putting our results into perspective with flavor bounds, direct detection, collider and atomic parity violation, we solidly conclude that weak scale right-handed neutrino stands for a plausible dark matter candidate in the presence of an early matter domination epoch.\\

\paragraph*{Acknowledgements\,:} 

FSQ and CS have been supported by the S\~{a}o Paulo Research Foundation (FAPESP) through Grant No 2015/15897-1. CS is supported by grant 2020/00320-9, S\~ao Paulo Research Foundation (FAPESP). FSQ acknowledges support from CNPq grants {\rm 303817/2018-6} and {\rm 421952/2018-0}, and ICTP-SAIFR FAPESP grant {\rm 2016/01343-7}, MEC, and UFRN. This work was supported by the Serrapilheira Institute (grant number Serra-1912-31613) and ANID – Millennium Program – $ICN2019_044$. The author JPN acknowledges the support from CAPES under the grant $88887.484542/2020-00$. 

\bibliographystyle{JHEPfixed.bst}
\bibliography{References.bib}%

\end{document}